%% file: sigconf.tex
\documentclass[sigconf]{acmart}
\AtBeginDocument{%
  }

\setcopyright{acmlicensed}
\copyrightyear{2018}
\acmYear{2018}
\acmDOI{XXXXXXX.XXXXXXX}
\acmConference[Conference acronym 'XX]{Make sure to enter the correct
  conference title from your rights confirmation email}{June 03--05,
  2018}{Woodstock, NY}
\acmISBN{978-1-4503-XXXX-X/2018/06}

\usepackage{xspace}
\usepackage[ruled,vlined,linesnumbered]{algorithm2e}
\usepackage[shortlabels]{enumitem}
\usepackage{listings}
\usepackage{xcolor}
\usepackage{colortbl}
\usepackage{tikz}
\usepackage{graphicx}
\usepackage{subcaption}
\usepackage[most]{tcolorbox}
\usepackage{wrapfig}
\usepackage{multirow}

\input{sections/defs}

\begin{document}

\title{Pre-QED: Environment Specifications for Agentic  Verification}

\author{Tristan Lim$^{*}$, Haoxin Tu$^{*}$, Mehtab Zafar$^{*}$, Zhengxiong Luo, Abhik Roychoudhury$^{\dagger}$}

\affiliation{%
  \institution{National University of Singapore, Singapore}
  \country{}
}
\email{\{tristanlimyirong{,} haoxin.tu{,} mehtab{,} luozx{,} abhik\}@nus.edu.sg}

\renewcommand{\shortauthors}{Lim, Tu, Zafar, Luo, Roychoudhury}
\renewcommand{\authors}{Tristan Lim, Haoxin Tu, Mehtab Zafar, Zhengxiong Luo, Abhik Roychoudhury}

\input{sections/abstract}


\makeatletter
\g@addto@macro\@translatedkeywords{%
  \par\medskip\noindent{\small $^{*}$Equal contribution with alphabetical order. $^{\dagger}$Corresponding author: Queries about the paper should be sent to the corresponding author \underline{abhik@nus.edu.sg}}.\par}
\makeatother


\maketitle

\input{sections/introduction}
\input{sections/background}
\input{sections/motivation}

\input{sections/approach}
\input{sections/stats}

\input{sections/technology}
\input{sections/related-work}
\input{sections/perspective}
\input{sections/data-ava}

\begin{acks}
This work was partially supported by a research project “AI for Program Reasoning”, project AI4SCH-2025-0084 under AI for Science program of National Research Foundation (NRF) Singapore. Any opinions, findings and conclusions, or recommendations expressed in this material are those of theauthor(s) and do not reflect the views of National Research Foundation, Singapore. Thanks to Anthropic's Frontier Red Team and Theorem for making the OSS-Verify proofs publicly available. 
\end{acks}

\balance
\bibliographystyle{ACM-Reference-Format}
\bibliography{references}




\end{document}

%% file: sections/defs.tex
\definecolor{mygreen}{rgb}{0,0.5,0}
\definecolor{myblue}{rgb}{0,0,1}
\definecolor{mymauve}{rgb}{0.58,0,0.82}
\definecolor{myblack}{rgb}{0.24,0.17,0.12}
\definecolor{awesome}{rgb}{1.0, 0.13, 0.32}
\definecolor{darkred}{rgb}{0.55, 0.0, 0.0}
\newcommand\code[1]{{\tt\small #1}}
\definecolor{dkgreen}{rgb}{0,0.6,0}
\definecolor{dkblue}{rgb}{0,0.4,0.4}
\definecolor{dkviolet}{rgb}{0.3,0,0.5}
\definecolor{dkred}{rgb}{0.6,0,0}
\DeclareRobustCommand{\stepnum}[1]{\tikz[baseline=-0.55ex]\path node[circle, fill=myblue, draw=none, text=white, inner sep=0.9pt, font=\tiny\bfseries]{#1};}
\lstdefinelanguage{Coq}{ 
    mathescape=true,
    texcl=false, 
    morekeywords=[1]{Section, Module, End, Require, Import, Export,
        Variable, Variables, Parameter, Parameters, Axiom, Hypothesis,
        Hypotheses, Notation, Local, Tactic, Reserved, Scope, Open, Close,
        Bind, Delimit, Definition, Let, Ltac, Fixpoint, CoFixpoint, Add,
        Morphism, Relation, Implicit, Arguments, Unset, Contextual,
        Strict, Prenex, Implicits, Inductive, CoInductive, Record,
        Structure, Canonical, Coercion, Context, Class, Global, Instance,
        Program, Infix, Theorem, Lemma, Corollary, Proposition, Fact,
        Remark, Example, Proof, Goal, Save, Qed, Defined, Hint, Resolve,
        Rewrite, View, Search, Show, Print, Printing, All, Eval, Check,
        Projections, inside, outside, Def},
    morekeywords=[2]{forall, exists, exists2, fun, fix, cofix, struct,
        match, with, end, as, in, return, let, if, is, then, else, for, of,
        nosimpl, when},
    morekeywords=[3]{Type, Prop, Set, true, false, option},
    morekeywords=[4]{pose, set, move, case, elim, apply, clear, hnf,
        intro, intros, generalize, rename, pattern, after, destruct,
        induction, using, refine, inversion, injection, rewrite, congr,
        unlock, compute, ring, field, fourier, replace, fold, unfold,
        change, cutrewrite, simpl, have, suff, wlog, suffices, without,
        loss, nat_norm, assert, cut, trivial, revert, bool_congr, nat_congr,
        symmetry, transitivity, auto, split, left, right, autorewrite},
    morekeywords=[5]{by, done, exact, reflexivity, tauto, romega, omega,
        assumption, solve, contradiction, discriminate},
    morekeywords=[6]{do, last, first, try, idtac, repeat},
    morecomment=[s]{(*}{*)},
    showstringspaces=false,
    morestring=[b]",
    morestring=[d]’,
    tabsize=3,
    extendedchars=false,
    sensitive=true,
    breaklines=false,
    basicstyle=\footnotesize,
    captionpos=b,
    columns=[l]flexible,
    identifierstyle={\ttfamily\color{black}},
    keywordstyle=[1]{\ttfamily\color{dkviolet}},
    keywordstyle=[2]{\ttfamily\color{dkgreen}},
    keywordstyle=[3]{\ttfamily\color{dkblue}},
    keywordstyle=[4]{\ttfamily\color{dkblue}},
    keywordstyle=[5]{\ttfamily\color{dkred}},
    stringstyle=\ttfamily,
    commentstyle={\ttfamily\color{dkgreen}},
    literate=
    {\\forall}{{\color{dkgreen}{$\forall\;$}}}1
    {\\exists}{{$\exists\;$}}1
    {<-}{{$\leftarrow\;$}}1
    {=>}{{$\Rightarrow\;$}}1
    {==}{{\code{==}\;}}1
    {==>}{{\code{==>}\;}}1
    {->}{{$\rightarrow\;$}}1
    {<->}{{$\leftrightarrow\;$}}1
    {<==}{{$\leq\;$}}1
    {\#}{{$^\star$}}1 
    {\\o}{{$\circ\;$}}1 
    {\@}{{$\cdot$}}1 
    {\/\\}{{$\wedge\;$}}1
    {\\\/}{{$\vee\;$}}1
    {++}{{\code{++}}}1
    {~}{{$\sim$}}1
    {\@\@}{{$@$}}1
    {\\mapsto}{{$\mapsto\;$}}1
    {\\hline}{{\rule{\linewidth}{0.5pt}}}1
}[keywords,comments,strings]

\newcommand{\autorocq}{\textsc{AutoRocq}\xspace} 
\newcommand{\rev}[1]{{#1}} %

\usepackage[most]{tcolorbox}
\newtcolorbox{promptresponsebox}[1][]{
  colback=white,               
  colframe=black,             
  fonttitle=\bfseries, 
  coltitle=black,              
  colbacktitle=gray!20,        
  title={Prompt-Response Example},
  boxrule=0.5pt,               
  top=1pt,                     
  bottom=1pt,                  
  left=4pt,                    
  right=4pt,                   
  toptitle=0.5pt,                
  bottomtitle=0.5pt,             
  #1
}
\theoremstyle{definition}
\newtheorem*{definition*}{Definition}
\usetikzlibrary{arrows.meta,positioning,shapes.geometric}

\newtcolorbox{resultbox}{
  colback=gray!10,     
  colframe=gray!70,    
  boxrule=0.8pt,       
  arc=4pt,             
  left=6pt, right=6pt, top=6pt, bottom=6pt, 
  enhanced
}



%% file: sections/abstract.tex
\begin{abstract}
LLM agents are making formal verification practical at scale. Recently, many attempts have raised this as an enticing possibility. However, formal verification at scale is only meaningful if deep formal specifications capturing functionality are available for checking. Unfortunately, recent works seek to bypass this issue by using shallow specifications devised for purposes other than verification. OSS-Verify from Anthropic and Theorem, for example, reuses the fuzz harnesses of open-source libraries as specifications and has an agent prove that each harness never crashes.
However, a fuzz harness follows a single sequence of API calls under a single calling context, so the resulting proof covers one usage scenario rather than the library. This is a core principle our team is well aware of, after more than a decade of experience in developing greybox fuzzing technology.

We argue that the bottleneck of agentic verification has shifted from \emph{how} to prove to \emph{what} to prove. Before zooming into verification at scale, we need deep environment specifications to prove, which is why we call our effort \emph{Pre-QED}. We propose \emph{verification harnesses}, executable harnesses that model all legal usages of an API, so that a proof about the harness is a proof about every well-behaved client.
To construct them, we use an agentic workflow that extracts the API's usage contract from code / documentation, and uses Linear-time Temporal Logic (LTL) to rule out illegal call orders.
We identify the challenges in proving verification harnesses and release our harnesses open-source as challenges for the community. The overarching vision is that verifying an open-source library should become as routine as fuzzing it, with a guarantee that covers every well-behaved client rather than a single scenario like a fuzzing harness.
\end{abstract}

%% file: sections/introduction.tex
\section{Introduction}
\label{sec:intro}

AI is making software vulnerabilities cheaper to find and to exploit, which raises the bar for defense: rather than adding more tests, we need guarantees that entire classes of bugs are absent~\cite{anthropic2026formalverification}.
Formal verification can provide such guarantees, but it has historically demanded years of expert effort; the seL4 microkernel, for instance, required roughly twenty person-years to verify~\cite{sel4}.
Large language model (LLM) agents are now changing this cost model by writing machine-checked proofs whose correctness is guaranteed by an independent proof checker rather than by the agent itself~\cite{autorocq}.
OSS-Verify~\cite{anthropic2026formalverification,theorem2026ossverifyproofs} takes this idea to open-source software at scale.
Instead of asking developers for specifications, it reuses the fuzz harnesses that projects already maintain for OSS-Fuzz~\cite{ossfuzz}, mechanically turns each harness into the theorem that the compiled program never crashes, and lets an agent prove that theorem in Rocq.
Applied to 239 entry points across 123 libraries, the pipeline produced kernel-checked proofs and uncovered hundreds of bugs along the way.

The price of this convenience is that the fuzz harness \emph{is} the specification, and a fuzz harness was never written to be one.
Its purpose is to drive the library efficiently with random bytes, so it typically hard-codes a single sequence of API calls under a single calling context.
A proof about the harness is therefore a proof about that one usage scenario, not about the library that other clients will call.
\autoref{sec:insufficient} makes this concrete with a handle-based API whose fuzz harness opens a handle, reads once, and closes it.
OSS-Verify can prove that this harness is crash-free for \emph{every} input up to 64\,MB, yet the library still contains an out-of-bounds read that any legal client triggers simply by reading twice from the same handle.
The proof is correct, but the guarantee it appears to give is not the one it actually gives.
As agents make proofs cheap, this gap becomes the dominant risk: a proof is only as trustworthy as the statement it establishes~\cite{agentic-ai-software-engineers}, and a ``verified'' label on a library whose theorem covers one call sequence can create a false sense of assurance.

Our position is that the bottleneck of agentic verification has shifted from \emph{how} to prove to \emph{what} to prove, and that the right object to prove is still a harness, but {\em not} a fuzz harness. Since our effort concerns what must be settled before a proof can be concluded, we call it \emph{Pre-QED}\footnote{QED stands for \emph{quod erat demonstrandum} (``which was to be demonstrated''), traditionally written at the end of a proof.}. We note that our research group has worked in fuzzing for a decade and our group has brought some of the earliest papers on greybox fuzzing, which have subsequently been adopted in popular fuzzers \cite{aflgo,aflfast}.

We propose \emph{verification harnesses}: executable harnesses that decode their input into an arbitrary sequence of API operations while admitting only those that respect the API contract, so that every input corresponds to a legal client and a proof about the harness is a proof about every well-behaved client (\autoref{sec:vharness}).
Because a verification harness keeps the familiar entry point and the same ``never reaches an error'' theorem, it plugs into the existing proof pipeline unchanged.
Building one requires meeting two requirements that mirror the gaps of fuzz harnesses: covering all valid API call sequences while excluding invalid orderings (control flow constraints), and generalizing calling contexts while enforcing preconditions across calls (data flow constraints).
We further observe that the ordering rules of an API are naturally temporal properties over call sequences, which lets us state them in Linear-time Temporal Logic (LTL)~\cite{pnueli1977ltl} and use them to guide and check an agentic workflow that extracts the usage contract from code and documentation and synthesizes the harness.

Building on these insights, we construct verification harnesses with an agentic workflow (\autoref{sec:vharness}).
A main agent first studies the library's code and documentation to extract its usage contract, separating ordering rules over the sequence of calls from data rules over the values passed between calls.
It then delegates two tracks: a harness subagent writes a candidate executable harness and refines it against code-coverage feedback, while an LTL subagent formalizes the ordering rules as temporal properties and refines them through spot checks and an LLM-as-judge assessment.
A final synthesis stage reconciles the two, so that the control flow of the harness implements the validated temporal properties and every guard is traceable to the rule it enforces.
On the motivating example, the resulting harness admits the client that reads twice from the same handle, so the proof attempt exposes the out-of-bounds read instead of missing it.
Proving such harnesses, however, is harder than proving fuzz harnesses: instead of following one fixed call sequence, the prover must reason about arbitrarily long sequences of API operations and about library state that persists across them (\autoref{sec:technology}).
We therefore release our verification harnesses as open challenges for agentic verifiers.

\paragraph*{Our contributions}
Our vision is that verifying an open-source library should become as routine as launching a fuzzing campaign, while the guarantee covers every well-behaved client rather than a single scenario.
Toward this vision, this paper makes the following contributions:
\begin{itemize}[leftmargin=*]
  \item We show that proofs derived from fuzz harnesses cover a single API usage scenario, and can hold while the library remains buggy. 
  \item We introduce \emph{verification harnesses}, which model all legal usages of an API, and an LTL-guided agentic workflow to construct them automatically. This also obviates the need to rely on 
  existing fuzzing harnesses for formal verification. We have an automated procedure for verification harness generation, and then we have checked them manually.
  \item We identify the challenges in proving verification harnesses and release our harnesses as open challenges, inviting the community to move agentic verification approaches from proving harnesses to proving the entire library. We have thus released the verification harnesses for the 29 libraries as open source at 
\textcolor{blue}{\url{https://pre-qed.github.io/}}.
\end{itemize}

%% file: sections/background.tex
\section{Background: The Oss-Verify Proof}
\label{sec:ossverify}

\subsection{Verification Pipeline}

OSS-Verify~\cite{anthropic2026formalverification}, developed jointly by Anthropic's Frontier Red Team and Theorem, demonstrates that proofs written by large language models (LLMs) can make formal verification practical for real-world open-source software at scale.
The work is motivated by the observation that AI has substantially lowered the cost of discovering and exploiting software vulnerabilities, which calls for guarantees that entire classes of bugs are absent, rather than for additional testing alone.
Formal verification can provide such guarantees, but it has traditionally required years of expert effort; for instance, the seL4 microkernel \cite{sel4} required approximately two person-years to build and twenty person-years to verify.
OSS-Verify reduces this cost in two ways: it reuses artifacts that developers already maintain, and it delegates proof construction to an LLM agent whose output is validated by an independent proof checker.

The pipeline takes as input an open-source library together with its existing fuzz harness, namely the \texttt{LLVMFuzzerTestOneInput} entry point already exercised by OSS-Fuzz~\cite{ossfuzz}.
No additional annotations or modifications by the library developers are required.
The library and the harness are compiled into a single statically linked RISC-V~\cite{riscv} executable.
Consistent with OSS-Fuzz practice, C/C++ code is built with AddressSanitizer (ASan)~\cite{asan} and UndefinedBehaviorSanitizer (UBSan)~\cite{ubsan}, so that invalid memory accesses and undefined behavior manifest as crashes rather than as silent corruption.
Reducing the verification target to one binary is essential for formal reasoning, because the behavior of the program is then fully determined by its machine code and by the formal models of the processor and the operating system on which it executes.

The executable is then combined with Theorem's formal semantics of RISC-V and of Linux processes, together with the property that the program never crashes, to generate a specification mechanically; the specification states, in essence, that the fuzz test always passes, and no part of it is written by hand.
A Claude agent subsequently constructs a proof of this specification in Rocq, which the Rocq kernel checks.
If the agent encounters a genuine bug, it patches the library, the specification is regenerated for the patched binary, and the agent proves the updated specification.
To discourage reward hacking, such as rewriting the program so that it trivially cannot crash, patches are penalized, additional credit is granted for proving which inputs a patch affects, and all patches are logged and audited.
Applied to 239 entry points across 123 OSS-Fuzz libraries, the pipeline uncovered hundreds of bugs, including some that fuzzing is unlikely to reach, such as defects triggered only by inputs of tens of megabytes.

\subsection{Proof Obligations}

The central premise of OSS-Verify is that no separate specification is authored: the fuzz harness itself defines what constitutes correct behavior for the target library.
This premise determines the scope of the verified claim.
The harness fixes the library API usage, so the proof covers exactly the calls issued by the harness rather than arbitrary client code.
The proven property is that, for every input smaller than 64\,MB, the program terminates normally, without memory-safety violations or assertion failures, under every modeled operating-system response and memory layout.
The 64\,MB bound is a parameter of the specification and can be raised by regenerating the specification and repeating the proof.
Since many harnesses additionally encode higher-level properties, such as a compress-then-decompress roundtrip, establishing that the harness never fails also establishes those properties.

The trustworthiness of the result rests on a small trusted computing base (TCB) consisting of three components.
First, the harness and the exact theorem statements must capture the intended notion of correctness, since the proof establishes only what they state.
Second, the environment model, i.e., the formal semantics of RISC-V and of single-threaded Linux processes, must faithfully reflect the behavior of real hardware and operating systems.
Third, the Rocq proof checker must be sound, together with the platform on which it runs, including the OCaml compiler, the operating system, and the hardware.
Given these components, the proofs themselves, which typically comprise approximately 600{,}000 lines of Rocq, require essentially no manual review.

The boundaries of this TCB also delineate the directions in which the approach must be extended: reducing the token cost of proof generation, reusing proofs across shared dependencies, and broadening the machine and operating-system models beyond statically linked, single-threaded RISC-V binaries, so that verification may become as routine as fuzzing. However, the work tries to use the fuzz harnesses from OSS-Fuzz as the program environment for verification, which may not be a good fit for verification purposes.

%% file: sections/motivation.tex
\section{Motivation: What the Proofs Leave Out}
\label{sec:insufficient}

\begin{figure}[t]
\begin{subfigure}{\columnwidth}
\begin{lstlisting}
typedef struct { const uint8_t *data; size_t cap, pos; } handle_t;

handle_t *h_open(const uint8_t *in, size_t n) { %*\label{line:h-open-start}*)
  if (n < 4) return NULL;
  handle_t *h = malloc(sizeof(handle_t));
  if (!h) return NULL;
  h->data = in; h->cap = n; h->pos = 0;
  return h;
} %*\label{line:h-open-end}*)
int h_read(handle_t *h, uint8_t *buf, size_t len) { %*\label{line:h-read-start}*)
  if (h->pos >= h->cap)
    return 0;      /* EOF: no remaining bytes */ %*\label{line:eof-check}*)
  if (len > h->cap)
    len = h->cap; /* BUG: clamps to h->cap, not h->cap-h->pos */ %*\label{line:bug-check}*)
  memcpy(buf,h->data+h->pos,len);/*OOB read on a second h_read()*/ %*\label{line:bug-memcpy}*)
  h->pos += len;
  return (int)len;
} %*\label{line:h-read-end}*)
void h_close(handle_t *h) { free(h); } %*\label{line:h-close}*)
\end{lstlisting}
\vspace{-2pt}
\caption{Target library code (\texttt{h\_open()}, \texttt{h\_read()}, \texttt{h\_close()}).}
\label{fig:fuzz-harness-lib}
\end{subfigure}

\vspace{4pt}

\begin{subfigure}{\columnwidth}
\begin{lstlisting}[firstnumber=18]
int FuzzHarness(const uint8_t *in, size_t n) { %*\label{line:fuzz-start}*)
  uint8_t buf[64];
  handle_t *h = h_open(in, n); %*\label{line:fuzz-call-start}*)
  if (h != NULL) {
    h_read(h, buf, sizeof(buf));
    h_close(h);
  } %*\label{line:fuzz-call-end}*)
  return 0;
} %*\label{line:fuzz-end}*)
\end{lstlisting}
\caption{Conventional fuzz harness exercising a single call sequence.}
\label{fig:fuzz-harness-driver}
\end{subfigure}
\caption{A representative library API with a stateful out-of-bounds read across multiple \texttt{h\_read()} calls (\subref{fig:fuzz-harness-lib}), and its conventional fuzz harness (\subref{fig:fuzz-harness-driver}).}
\vspace{-1em}
\label{fig:fuzz-harness-example}
\end{figure}

\subsection{Motivating Example}

\input{figures/gap}

\input{figures/construction-workflow}

To illustrate why a proof about a fuzz harness falls short of verifying the underlying library, we show a representative library API (\autoref{fig:fuzz-harness-lib}) and its conventional fuzz harness (\autoref{fig:fuzz-harness-driver}).

The library exposes three functions: \texttt{h\_open()} initializes a handle \texttt{h} over an input buffer \texttt{in} of length \texttt{n} with capacity \texttt{h->cap = n} and cursor \texttt{h->pos = 0} (lines~\ref{line:h-open-start}--\ref{line:h-open-end}); \texttt{h\_read()} copies up to \texttt{len} bytes from \texttt{h->data + h->pos} into a caller buffer \texttt{buf}, clamping the read at stream end and advancing \texttt{h->pos} (lines~\ref{line:h-read-start}--\ref{line:h-read-end}); and \texttt{h\_close()} frees \texttt{h} (line~\ref{line:h-close}).
Following standard OSS-Fuzz practice, the developer-written \texttt{FuzzHarness} (lines~\ref{line:fuzz-start}--\ref{line:fuzz-end}) opens a handle \texttt{h} on the fuzzer input, performs a single \texttt{h\_read()} into a 64-byte buffer \texttt{buf} with \texttt{len = sizeof(buf)}, and closes \texttt{h}.

\smallskip
\textbf{What the Proof Establishes.}
When \texttt{FuzzHarness} is fed into the OSS-Verify pipeline (\autoref{sec:ossverify}), verification passes: the prover successfully establishes a machine-checked proof that the compiled binary is crash-free (and trace-equivalent across builds) for \emph{all} inputs up to 64\,MB.
Indeed, along the single execution path fixed by lines~\ref{line:fuzz-call-start}--\ref{line:fuzz-call-end}, \texttt{h\_read()} is invoked at most once per handle with \texttt{h->pos == 0} and \texttt{len == 64}.
When \texttt{h->pos == 0}, clamping \texttt{len} to \texttt{h->cap} at line~\ref{line:bug-check} is exact: whether \texttt{n < 64} or \texttt{n >= 64}, the \texttt{memcpy} at line~\ref{line:bug-memcpy} copies $\min(64, \texttt{n})$ bytes from \texttt{h->data + 0} and stays within bounds, so the verification goal holds unconditionally.

\smallskip
\textbf{What the Proof Misses.}
Although formal verification passes on \texttt{FuzzHarness}, the library still contains a genuine stateful out-of-bounds read at line~\ref{line:bug-check}.
Like standard stream APIs (e.g., \texttt{fread} or \texttt{gzread}), \texttt{h\_read()} is designed to be called repeatedly in chunks until it returns \texttt{0} at EOF (\texttt{h->pos >= h->cap}, line~\ref{line:eof-check}); clamping a caller's request to the remaining bytes is the library's own responsibility.
However, line~\ref{line:bug-check} clamps \texttt{len} against the total capacity \texttt{h->cap} rather than the remaining bytes \texttt{h->cap - h->pos}.
Consequently, any legal client that reads in two chunks—for instance, calling \texttt{h\_read()} with \texttt{len = 40} twice on a stream of \texttt{h->cap = 64} bytes—sees \texttt{h->pos = 40 < h->cap} (not EOF) and \texttt{len = 40 <= h->cap} on the second call, causing \texttt{memcpy} at line~\ref{line:bug-memcpy} to read 16 bytes past the end of \texttt{h->data}.
Because \texttt{FuzzHarness} hard-codes a single one-shot sequence ($\texttt{h\_open()} \rightarrow \texttt{h\_read()} \rightarrow \texttt{h\_close()}$), the verified theorem is completely silent on chunked reads.

\subsection{Gap Analysis}
\label{sec:gaps}

To understand why this gap is structural rather than incidental to a single harness, we model the legal usages of a library API as an \emph{automaton} (\autoref{fig:api-automaton}).
It is a finite automaton over finite sequences of API calls, in which every state except \textsf{Misuse} is accepting: a call sequence is legal if and only if it never reaches \textsf{Misuse}, so every prefix of a legal sequence is itself legal.
This choice follows from the fact that all ordering rules we consider are safety properties: every violation is witnessed by a finite \emph{bad prefix}~\cite{kupferman2001model}, which here is a sequence of calls that reaches \textsf{Misuse}, and a harness enforces these rules by never taking a transition into \textsf{Misuse}.
Liveness rules, such as requiring that every opened handle is eventually closed, cannot be witnessed by a finite prefix and would instead require a B\"uchi automaton \cite{vardi1986automata} over infinite sequences (\autoref{sec:technology}).

In this automaton, states represent the lifecycle of library objects—uninitialized or closed (\texttt{NONE}), freshly opened at \texttt{h->pos = 0} (\texttt{OPEN}), and actively being read (\texttt{READ})—transitions correspond to API calls, and each transition is parameterized by a \emph{calling context} (the arguments passed to the call and the internal state accumulated along the path, such as \texttt{len}, \texttt{h->cap}, and \texttt{h->pos}).
Viewed through this lens, repurposing a conventional fuzz harness as a formal specification suffers from two structural limitations:

\textbf{(1) Sequence (Single Path vs.\ Full Automaton).}
In \autoref{fig:fuzz-harness-driver}, \texttt{FuzzHarness} traverses only the three-step subgraph \stepnum{1}\,\texttt{h\_open()} $\rightarrow$ \stepnum{2}\,\texttt{h\_read()} $\rightarrow$ \stepnum{3}\,\texttt{h\_close()} (\texttt{NONE} $\rightarrow$ \texttt{OPEN} $\rightarrow$ \texttt{READ} $\rightarrow$ \texttt{NONE}) and terminates.
Every other valid cycle in \autoref{fig:api-automaton}, including the black dashed self-loop on \texttt{READ} that invokes \texttt{h\_read()} repeatedly and the reopen cycle across handles, is syntactically absent from the binary and therefore invisible to the theorem.
A natural attempt to cover more of the automaton is to invoke APIs in random order; however, unconstrained random sequences immediately take illegal transitions—such as calling \texttt{h\_read()} or \texttt{h\_close()} in state \texttt{NONE} (the gray dotted edge to \textsf{Misuse} in \autoref{fig:api-automaton})—and produce client-side misuse crashes that falsify the theorem without implicating the library.
To verify the full automaton without admitting such misuse, the harness must explicitly encode the boundary between legal and illegal call sequences.

\textbf{(2) API Call Context (Single Scenario vs.\ General Preconditions).}
Even along the transitions it does take, a fuzz harness hard-codes a single calling scenario to satisfy API preconditions, rather than ranging over all valid contexts.
In \autoref{fig:fuzz-harness-driver}, the transition from \texttt{OPEN} to \texttt{READ} is always taken at \texttt{h->pos == 0} with a fixed request size \texttt{len = 64} on a single handle \texttt{h}.
In reality, the precondition of \texttt{h\_read()} permits \emph{any} \texttt{len <= sizeof(buf)} at \emph{any} accumulated cursor offset \texttt{h->pos} produced by prior reads in state \texttt{READ}, as well as multiple handles coexisting in memory.
By collapsing this space of legal contexts to a single point (\texttt{len = 64}, \texttt{h->pos = 0}), the proof in \autoref{sec:insufficient} passes for that one scenario while leaving genuine library defects triggered under general contexts (\autoref{fig:fuzz-harness-lib}, line~\ref{line:bug-check}) unverified.

%% file: figures/gap.tex
\begin{figure}[t]
\centering
\resizebox{\columnwidth}{!}{%
\begin{tikzpicture}[
  >=Stealth,
  state/.style={circle, draw=black, thick, minimum size=0.88cm, font=\footnotesize\bfseries, fill=white},
  errstate/.style={circle, draw=gray!65!black, thick, dotted, minimum size=0.88cm, font=\scriptsize\bfseries, text=gray!65!black, fill=gray!8},
  fuzz/.style={->, very thick, myblue},
  missed/.style={->, very thick, black, dashed},
  illegal/.style={->, thick, gray!65!black, densely dotted},
  stepbadge/.style={circle, fill=myblue, text=white, inner sep=0.9pt, font=\tiny\bfseries}
]
  \draw[rounded corners=6pt, fill=blue!4, draw=myblue!45, dashed, thin]
    (1.92, -1.42) rectangle (9.02, 0.76);
  \node[anchor=north east, font=\tiny\bfseries, text=myblue] at (8.98, -1.46) {Fuzz harness subgraph};

  \node[errstate]  (err)  at (0.0,  0) {Misuse};
  \node[state]     (none) at (2.55, 0) {\texttt{NONE}};
  \node[state]     (open) at (5.40, 0) {\texttt{OPEN}};
  \node[state]     (read) at (8.40, 0) {\texttt{READ}};

  \node[rounded corners=2pt, fill=black, text=white, inner sep=2pt, font=\tiny\bfseries] (startpt) at (2.55, -1.72) {START};
  \draw[->, very thick] (startpt.north) -- (none.south);

  \draw[illegal] (none) --
    node[above=2pt, font=\scriptsize, text=gray!65!black] {\texttt{h\_read()}}
    node[below=2pt, font=\scriptsize, text=gray!65!black] {\texttt{h\_close()}} (err);

  \draw[missed, in=118, out=62, looseness=6.5] (none) to
    node[above=2.5pt, font=\scriptsize, text=black] {\texttt{h\_open()=NULL}} (none);

  \draw[fuzz] (none) -- (open);
  \node[above=2pt, font=\scriptsize, text=myblue] at (3.975, 0) {\stepnum{1}\,\texttt{h\_open()!=NULL}};

  \draw[missed, bend left=30] (open.-145) to
    node[below=0.5pt, pos=0.5, font=\scriptsize, text=black] {\rev{\texttt{h\_close()}}} (none.-30);

  \draw[fuzz] (open) -- (read);
  \node[above=2pt, font=\scriptsize, text=myblue] at (6.90, 0) {\stepnum{2}\,\texttt{h\_read()}};
  \node[below=2pt, font=\tiny, text=myblue] at (6.90, 0) {(\texttt{len=64,\ h->pos=0})};

  \draw[fuzz, out=-132, in=-48, looseness=0.85] (read) to coordinate[pos=0.5] (c3) (none);
  \node[above=2pt, font=\scriptsize, text=myblue] at (c3) {\stepnum{3}\,\texttt{h\_close()}};

  \draw[missed, in=118, out=62, looseness=6.5] (read) to
    node[above=2.5pt, font=\scriptsize, text=black, align=center] (missedlbl) {
      \texttt{h\_read()}\\
      {\tiny (\texttt{len<=64,\ h->pos>0})}
    } (read);
  \node[rounded corners=2pt, fill=red!12, draw=darkred, solid, thick, inner xsep=3.5pt, inner ysep=1.2pt, font=\tiny\bfseries, text=darkred, anchor=south] at ([yshift=1.5pt]missedlbl.north) {BUG: triggers OOB read};
\end{tikzpicture}%
}
\vspace{2pt}
\resizebox{\columnwidth}{!}{%
  {\scriptsize\bfseries
    \tikz[baseline=-0.5ex,>=Stealth]\draw[->, very thick, myblue] (0,0) -- (0.48,0);~\textcolor{myblue}{Fuzz harness path (\stepnum{1}$\to$\stepnum{2}$\to$\stepnum{3})}\quad
    \tikz[baseline=-0.5ex,>=Stealth]\draw[->, very thick, black, dashed] (0,0) -- (0.48,0);~\textcolor{black}{Missed legal paths / contexts}\quad
    \tikz[baseline=-0.5ex,>=Stealth]\draw[->, thick, gray!65!black, densely dotted] (0,0) -- (0.48,0);~\textcolor{gray!65!black}{Illegal misuse (ignored)}%
  }%
}
\caption{\rev{Finite automaton (over finite sequences of API calls)} of the library API. \rev{\texttt{NONE} is the initial state, and all states except \textsf{Misuse} are accepting, so a call sequence is legal iff it never reaches \textsf{Misuse}.} The conventional fuzz harness (\autoref{fig:fuzz-harness-driver}) covers only the shaded blue subgraph (\stepnum{1}$\to$\stepnum{2}$\to$\stepnum{3}) under a single context (e.g., \texttt{len = 64}, \texttt{h->pos = 0}), missing the black dashed self-loop on \texttt{READ} for repeated \texttt{h\_read()} calls that triggers the OOB read.}
\vspace{-1em}
\label{fig:api-automaton}
\end{figure}
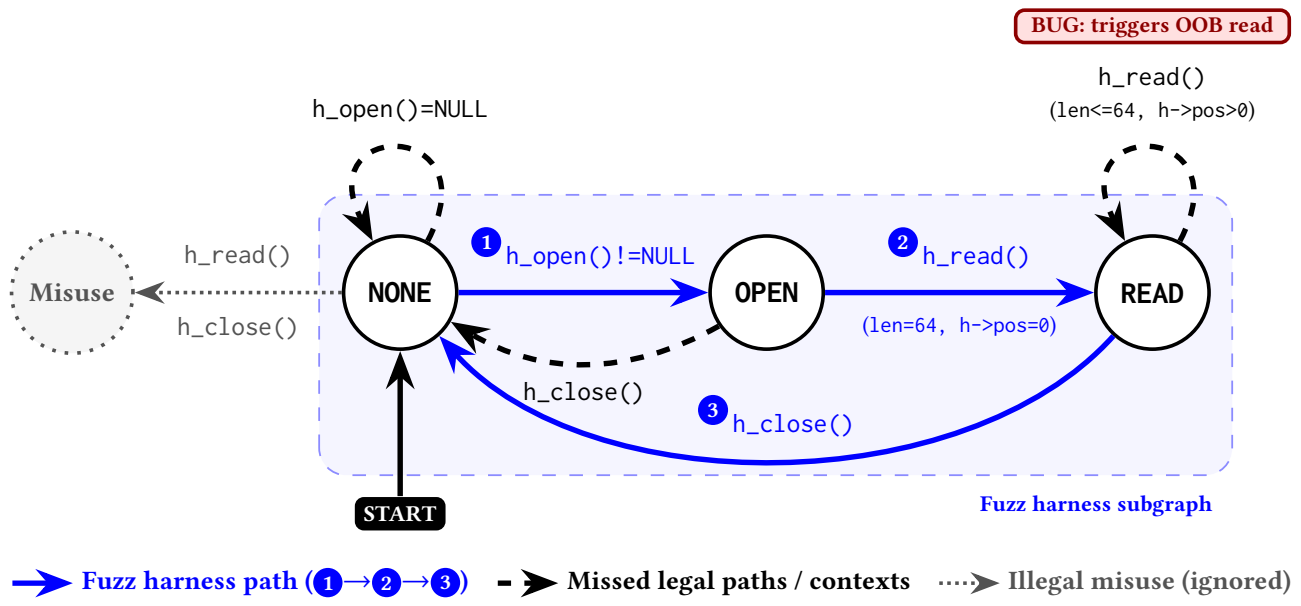

%% file: figures/construction-workflow.tex
\begin{figure*}[t]
  \centering
  \includegraphics[width=\textwidth]{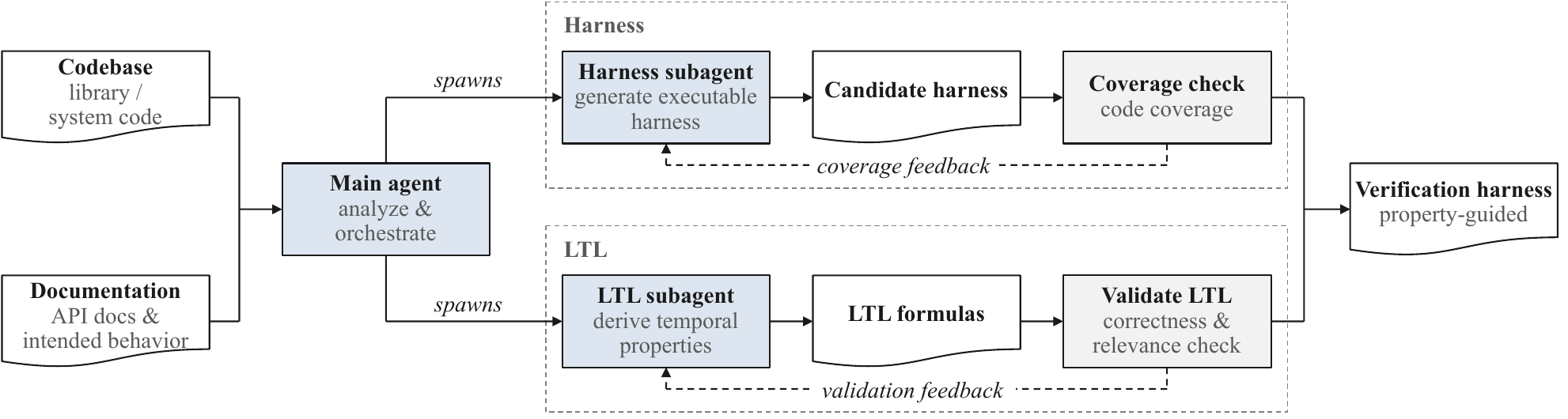}
  \caption{Agentic verification-harness construction workflow.}
  \label{fig:agentic-veri-harness-workflow}
\end{figure*}

%% file: sections/approach.tex
\section{Construction of Verification Harnesses}
\label{sec:vharness}

\subsection{Definition and Requirements}

The proofs discussed in \autoref{sec:ossverify} are only as general as the fuzz harness from which their specification is derived (\autoref{sec:insufficient}).
Because that harness (\autoref{fig:fuzz-harness-driver}) fixes a single API call sequence and a single calling context, the resulting guarantee covers one usage scenario rather than the library as a whole.
This is why the proof in \autoref{sec:insufficient} is silent on the stateful out-of-bounds read of \autoref{fig:fuzz-harness-lib}, which manifests only when \texttt{h\_read()} is invoked repeatedly on the same handle.
We therefore introduce the notion of a \emph{verification harness}: an executable harness that models \emph{all} legal usages of the target library's API, so that a proof about the harness is a proof about every well-behaved client.
\autoref{fig:harness-example} illustrates the idea for the target library in \autoref{fig:fuzz-harness-lib}.
Instead of issuing a fixed sequence of calls, the harness decodes its input bytes into an arbitrary sequence of API operations, and it admits only those operations that respect the API contract, so that every decoded input corresponds to a legal client.
The theorem to be proved is then that, for all inputs, the program never reaches an error state.

Two requirements follow directly from the gaps identified in \autoref{sec:insufficient}.
\begin{enumerate}[leftmargin=*]
  \item \textbf{Control flow (API call sequence):} The harness must cover every valid sequence of API calls while excluding invalid call orders. In \autoref{fig:harness-example}, this is the role of the explicit state variable that records whether a handle is open and thereby encodes the ordering rules (O1, O2).
  \item \textbf{Data flow (API call context):} The harness must generalize the calling contexts, rather than fixing a single scenario, while enforcing the API preconditions across calls. In \autoref{fig:harness-example}, the handle, the buffer, and its length are derived only from values that the harness owns, which satisfies the data rules (D1--D3), whereas the remaining inputs, such as the raw bytes passed to the open operation, remain unconstrained.
\end{enumerate}
A harness that violates the first requirement either misses reachable behavior or exercises illegal usages whose failures are attributable to the client rather than to the library; a harness that violates the second requirement either proves too little, by remaining tied to one calling scenario, or too much, by admitting contexts that violate the preconditions.

\subsection{Verification Harness Generation}


Writing such a harness by hand requires a thorough understanding of the library's API contract, and the contract is typically scattered across the source code and the documentation.
We therefore construct verification harnesses with an agentic workflow (\autoref{fig:agentic-veri-harness-workflow}) and model the required control-flow constraints with Linear-time Temporal Logic (LTL)~\cite{pnueli1977ltl}.
LTL is a natural fit because the ordering rules of an API, such as ``a handle is read only after it has been opened and before it is closed'', are properties over sequences of calls.
Expressing them as LTL properties makes the intended usage protocol explicit and checkable, and it provides a precise reference against which the generated harness can be assessed.

The workflow proceeds in four stages.
First, a main agent \emph{analyzes and orchestrates}: it studies the codebase together with the API documentation to understand the library and its intended behavior, and then delegates two focused tracks.
Second, a \emph{harness subagent} generates a candidate executable harness; the candidate is validated through a code-coverage check, and the resulting feedback is returned to the subagent for refinement.
Third, an \emph{LTL subagent} derives temporal properties from the code and the documentation; these formulas are validated through spot checks and an LLM-as-judge assessment of their correctness and relevance, and the feedback is again used to refine them.
Finally, the \emph{synthesis} stage combines the validated outputs into the final verification harness, which is both executable and guided by the validated properties.

\begin{figure}[t]
\input{assets/code}
\caption{An example verification harness for the library in \autoref{fig:fuzz-harness-lib}, generalizing the fuzz harness in \autoref{fig:fuzz-harness-driver}.}
\vspace{-1em}
\label{fig:harness-example}
\end{figure}

\subsection{An Example of Verification Harness}

We use the harness in \autoref{fig:harness-example} to illustrate how the workflow in \autoref{fig:agentic-veri-harness-workflow} produces a verification harness for the target library in \autoref{fig:fuzz-harness-lib} (contrasting with its conventional fuzz harness in \autoref{fig:fuzz-harness-driver}).
The target is a small library with a handle-based API consisting of \texttt{h\_open()}, \texttt{h\_read()}, and \texttt{h\_close()}.
Throughout the example, O1 and O2 denote the API’s ordering rules, and D1–D3 denote its data rules. The workflow extracts these rules as follows.

\smallskip
{\bf Analyze \& Orchestrate.}
The main agent reads the library's code and documentation, and it extracts the usage contract of the API.
Ordering rules: (O1) a read is legal only while a handle is open; (O2) at most one handle is live at a time, and a close is legal only for a handle that is currently open.
Data rules: (D1) a handle is usable only if \texttt{h\_open()} returned it as non-\texttt{NULL};
(e.g., \texttt{requires \textbackslash valid(h)}) (D2) after \texttt{h\_close()}, the handle must not be used again; (D3) the length passed to \texttt{h\_read()} must not exceed the size of the buffer passed alongside it. (e.g., \texttt{requires \textbackslash valid(buf + (0 .. len-1))})
The main agent then delegates the ordering rules to the LTL track and the construction of an executable harness to the harness track.

\smallskip
{\bf LTL Subagent.}
From the ordering rules, the LTL subagent derives temporal properties over the sequence of API calls, for example
\begin{align*}
  \varphi_{\mathrm{O1}} &= \neg\,\mathit{read}\ \mathbf{W}\ \mathit{open} \;\wedge\; \mathbf{G}\big(\mathit{close} \rightarrow \mathbf{X}(\neg\,\mathit{read}\ \mathbf{W}\ \mathit{open})\big),\\
  \varphi_{\mathrm{O2}} &= \mathbf{G}\big(\mathit{open} \rightarrow \mathbf{X}(\neg\,\mathit{open}\ \mathbf{W}\ \mathit{close})\big) \;\wedge\; \big(\neg\,\mathit{close}\ \mathbf{W}\ \mathit{open}\big),
\end{align*}
where $\mathbf{G}$, $\mathbf{X}$, and $\mathbf{W}$ denote ``always'', ``next'', and ``weak until''.
In words, a read never occurs before the first open or after a close until the next open, and a handle is never opened twice without an intervening close, nor closed unless it is open.
Spot checks and the LLM-as-judge assessment confirm that these formulas agree with the documentation and that none is vacuous or irrelevant to the API.
The data rules D1--D3 are constraints on values rather than on the order of calls, so they are not expressed in LTL; they are handled by the harness track.

\smallskip
{\bf Harness Subagent.}
The harness subagent turns the contract into an executable harness (\texttt{VerificationHarness} in \autoref{fig:harness-example}).
To cover \emph{all} valid call sequences, it replaces the fixed sequence of calls in \texttt{FuzzHarness} (\autoref{fig:fuzz-harness-driver}) with a decoder: each pair of input bytes selects one operation (\texttt{kind \% 3}) and its argument, and the loop continues until the input is exhausted.
Every sequence of operations is thereby reachable from some input, including the repeated reads that the fixed sequence of \autoref{fig:fuzz-harness-driver} never issues.
To restrict the sequences to legal ones, it realizes the automaton of $\varphi_{\mathrm{O1}}$ and $\varphi_{\mathrm{O2}}$ as the state variable \texttt{st}.
A read or a close whose guard does not hold (\texttt{st != OPEN}) is skipped, so that the input remains legal and no input is rejected, and an open in the \texttt{OPEN} state first closes the live handle, which preserves O2.
For the data rules, the harness owns the only handle \texttt{h} and the only buffer \texttt{buf}, updates \texttt{st} from the return value of \texttt{h\_open()} (D1), resets \texttt{h} on close (D2), and derives the read length as \texttt{arg \% sizeof(buf)} (D3).
Inputs that the contract leaves open remain unconstrained: the raw bytes passed to \texttt{h\_open()} are forwarded without validation, because rejecting malformed inputs is the library's responsibility, and constraining them would exclude exactly the behavior that the proof is meant to cover.
The coverage check then confirms that every API function and every guard branch of the harness is reached; unreached branches are returned to the subagent as feedback, for instance to add a missing reopen case after a close.

\smallskip
{\bf Synthesis.}
Finally, the synthesis stage checks that the candidate harness and the LTL formulas agree, and it emits the final verification harness.
The check is traceable by construction, since each guard and state update in the harness is annotated with the rule (O1, O2, D1--D3) that it enforces.
The synthesized harness additionally closes any handle that is still open at the end of the input, so that every run ends in a state permitted by the rules and the run does not leak.
The resulting harness is executable, as a fuzz harness is, and property-guided, because its control flow implements the validated temporal properties.
The theorem to be proved is that, for all inputs, the program never reaches an error, which by construction is a statement about all legal clients of the API.
Unlike the proof for the fuzz harness in \autoref{sec:insufficient}, this theorem cannot be established for the library of \autoref{fig:fuzz-harness-lib}: an input that decodes to \texttt{h\_open()} on a 64-byte stream followed by two \texttt{h\_read()} operations of 40 bytes each leaves \texttt{len} unclamped at line~\ref{line:bug-check} on the second call (\texttt{40 <= 64}) and reads 16 bytes past the end of the buffer at line~\ref{line:bug-memcpy}, so the proof attempt exposes the defect rather than missing it.

%% file: assets/code.tex
\begin{lstlisting}
/* Verification harness: every decoded input is a legal
   client of the API. Theorem to prove: for all inputs,
   this program never reaches an error. */

/* Call-order state: is a handle open? (O1, O2) */
enum { NONE, OPEN } st = NONE;
/* The open handle, or NULL if none (D1, D2) */
handle_t *h   = NULL;
/* Read buffer of known size, used to bound len (D3) */
uint8_t   buf[64];

int VerificationHarness(const uint8_t *in, size_t n) {
  size_t i = 0;
  while (n - i >= 2) {
    uint8_t kind = in[i], arg = in[i + 1];
    i += 2;
    switch (kind % 3) {

    case 0: /* OPEN */
      /* O2: at most one live handle, so close it first */
      if (st == OPEN) { h_close(h); st = NONE; }
      if (arg > n - i) break;
      /* Data left unconstrained: the library must check it */
      h  = h_open(in + i, arg);
      i += arg;
      /* D1: the handle is usable only if it is non-NULL */
      st = (h != NULL) ? OPEN : NONE;
      break;

    case 1: /* READ */
      /* O1, D1: read only an open handle; else skip */
      if (st != OPEN) break;
      /* D3: len < sizeof(buf), so the read fits in buf */
      h_read(h, buf, arg % sizeof(buf));
      break;

    case 2: /* CLOSE */
      /* O2: close only an open handle; else skip */
      if (st != OPEN) break;
      h_close(h);
      /* D2: drop the closed handle so it is never reused */
      h  = NULL;
      st = NONE;
      break;
    }
  }
  /* Close a handle left open at the end, so nothing leaks */
  if (st == OPEN) h_close(h);
  return 0;
}
\end{lstlisting}

%% file: sections/stats.tex
\section{Statistics of Verification Harnesses}
\label{sec:stats}

\input{figures/stats-info.tex}

\textbf{Subjects.}
Of the libraries verified by OSS-Verify (\autoref{sec:ossverify}), 32 have publicly released proofs\footnote{\url{https://github.com/theorem-labs/oss-verify-proofs/tree/main\#what-is-proved}}, and we take these 32 as our candidate subjects.
We construct verification harnesses for 29 of them and exclude the remaining three for the following reasons.
One subject {\tt reac-pw} is an application rather than a library: it has no API that clients call in sequence, so the legal call sequences that a verification harness should model are not well defined.
Two subjects, {\tt openssl} and {\tt mbedtls}, each expose more than 2,000 API functions; a harness that covers such an API would be too large to be reviewed properly, and an unreviewed harness may not serve as a trusted specification.

\smallskip
\textbf{Comparison with Fuzzing Harnesses.}
\autoref{tab:harness-comparison} compares, for each of the 29 subjects, the existing fuzz harness used by OSS-Verify with our verification harness.
The subjects are ordered by increasing size (LoC) of the core library code, excluding the tests, fuzzers, and examples.
For both harnesses, we report the API coverage (Cov. \%), i.e., the number of APIs that appear in the harness divided by the total number of APIs in the library, and for the verification harness we additionally report the number of LTL properties (\# LTL) extracted by the LTL subagent (\autoref{sec:vharness}).

On average, the verification harnesses reach 89.3\% coverage, compared with 22.1\% for the fuzz harnesses, an improvement of 67.2 percentage points.
Two subjects, {\tt reac-repacer} and {\tt cgif}, are already fully covered by their fuzz harnesses and remain at 100\%, and the verification harnesses reach 100\% coverage on 11 subjects in total, compared with two for the fuzz harnesses.
The largest gains occur for {\tt plugin-hostd} (0.0\% to 100.0\%), {\tt pupnp} (0.8\% to 99.6\%), {\tt cjson} (6.5\% to 100.0\%), and {\tt tinyobjloader} (5.5\% to 96.4\%), whose fuzz harnesses cover less than 10\% of the API, whereas the verification harnesses invoke nearly all API functions in arbitrary legal orders.
In total, the workflow extracts 234 LTL properties, 8.1 per subject on average (ranging from 0 to 33, the latter for {\tt libreac}), and each of them is enforced by the control flow of the corresponding harness.
Two subjects, {\tt reac-repacer} and {\tt fast\_float}, have no LTL properties because their APIs are stateless: API calls do not carry anything between their calls, so there is no ordering constraint on the call sequence to express.
Cov. measures the breadth of the API that a harness exercises; it does not capture call sequences or calling contexts, whose generality follows instead from the construction of the verification harness (the input decoder together with the LTL-guided guards in \autoref{sec:vharness}).
Accordingly, a fuzz harness with full API coverage, such as those of {\tt reac-repacer} and {\tt cgif}, may still fix a single call sequence and calling context.
Even by this breadth measure alone, however, the numbers indicate that the theorems obtained from fuzz harnesses leave most of the API of the libraries unexercised, and the verification harnesses close much of this gap.
Where the verification harness does not reach every API, we check the uncovered APIs, and the gaps fall into a few categories: APIs that only exist in other build configurations (e.g., {\tt libpcap}, {\tt pupnp}), and APIs that spawn processes or touch the filesystem (e.g., {\tt leptonica}), code that is only used by a daemon or that depends on private libraries (e.g., {\tt varnish}, {\tt omx-clap-core}).

\smallskip
{\bf Manual Efforts Involved.}
Our workflow produces the verification harnesses automatically, and we spend manual effort mainly to further increase our confidence in them before we use them as specifications.
Because each harness is short and structured around the extracted API usage protocol, we can review it easily: we mostly look at code quality (e.g., unnecessary complexity added to increase code coverage, unexplained use of constants) and at whether the harness has the desired shape, i.e., that it drives the API through the legal call sequences and calling contexts described in \autoref{sec:gaps} instead of a single hard-coded scenario.
We also carefully check the library documentation to make sure that the API call sequences encoded in each harness align with the sequences that the documentation expects.
In all cases, we need only light-weight inspection and do not have to rewrite the harnesses.

%% file: figures/stats-info.tex
\definecolor{covgreen}{HTML}{1A7F37}
\newlength{\covw}\settowidth{\covw}{\small Fuzz harness}\addtolength{\covw}{1pt}
\newcolumntype{C}[1]{>{\centering\arraybackslash}p{#1}}
\newcommand{\cov}[1]{\pgfmathsetmacro{\covtint}{5 + 0.70*(#1)}\edef\covspec{covgreen!\covtint!white}\expandafter\cellcolor\expandafter{\covspec}#1}

\begin{table}[t]
\centering
\caption{Comparison between fuzzing and verification harnesses in terms of API usage coverage (refs. {\bf API Cov(\%)}) and the number of LTL constraints (refs. {\bf \# LTL}), ordered by increasing size (LoC) of the core library code.}
\label{tab:harness-comparison}
\small
\setlength{\tabcolsep}{5pt}
\begin{tabular}{lrC{\covw}C{\covw}r}
\toprule
& & Fuzz harness & \multicolumn{2}{c}{Verification harness} \\
\cmidrule(lr){3-3}\cmidrule(lr){4-5}
Subject & Size & Cov. (\%) & Cov. (\%) & \# LTL \\
\midrule
reac-repacer & 290 & \cov{100.0} & \cov{100.0} & 0 \\
libdiginet & 1,659 & \cov{85.1} & \cov{100.0} & 5 \\
cgif & 1,760 & \cov{100.0} & \cov{100.0} & 2 \\
yajl-ruby & 3,087 & \cov{11.5} & \cov{100.0} & 3 \\
double-conversion & 3,921 & \cov{6.2} & \cov{84.4} & 3 \\
fast\_float & 3,945 & \cov{8.3} & \cov{100.0} & 0 \\
cjson & 4,002 & \cov{6.5} & \cov{100.0} & 5 \\
bzip2 & 5,871 & \cov{33.3} & \cov{100.0} & 5 \\
json-c & 6,198 & \cov{31.5} & \cov{78.5} & 8 \\
skcms & 7,095 & \cov{10.5} & \cov{100.0} & 2 \\
plugin-hostd & 9,726 & \cov{0.0} & \cov{100.0} & 1 \\
coturn & 11,938 & \cov{37.6} & \cov{97.3} & 6 \\
tinyobjloader & 13,136 & \cov{5.5} & \cov{96.4} & 4 \\
libreac & 13,202 & \cov{17.9} & \cov{100.0} & 33 \\
zlib & 17,445 & \cov{58.0} & \cov{93.2} & 6 \\
speex & 18,863 & \cov{35.8} & \cov{100.0} & 6 \\
omx-clap-core & 20,378 & \cov{0.0} & \cov{76.5} & 7 \\
boost-json & 26,496 & \cov{5.1} & \cov{63.3} & 4 \\
pupnp & 34,463 & \cov{0.8} & \cov{99.6} & 20 \\
simdjson & 36,709 & \cov{13.1} & \cov{90.3} & 15 \\
boost-beast & 39,776 & \cov{2.6} & \cov{62.3} & 16 \\
libpcap & 46,453 & \cov{9.4} & \cov{88.7} & 4 \\
zstd & 50,891 & \cov{30.3} & \cov{90.4} & 13 \\
msgpack-c & 76,964 & \cov{2.1} & \cov{82.4} & 11 \\
glaze & 85,456 & \cov{7.9} & \cov{92.5} & 5 \\
jsoncons & 89,201 & \cov{4.2} & \cov{61.0} & 10 \\
spirv-tools & 109,870 & \cov{6.0} & \cov{62.9} & 18 \\
varnish & 120,139 & \cov{0.7} & \cov{71.7} & 20 \\
leptonica & 144,883 & \cov{10.2} & \cov{97.5} & 2 \\
\bottomrule
\end{tabular}
\vspace{-0.5em}
\end{table}

%% file: sections/technology.tex
\section{Towards Proving Verification Harnesses}
\label{sec:technology}

\textbf{Proving Lemmas over Complex API Call Sequences.}
The existing entry-point proofs succeed largely because a fuzz harness follows a single, fixed sequence of API calls, each starting from freshly initialized library state.
A verification harness removes this structure: its decoding loop admits arbitrarily long sequences of operations, and the library state, such as the cursor \texttt{h->pos} in \autoref{fig:fuzz-harness-lib}, persists and evolves across them.
Following each sequence symbolically is no longer possible, so the prover must instead establish a loop invariant that characterizes every library state reachable by a legal client, together with per-API lemmas showing that each operation, started in a state satisfying the invariant and called with arguments that satisfy its preconditions, terminates without error and preserves the invariant.
For the example library, the invariant must state that \texttt{h->pos} never exceeds \texttt{h->cap}, which is exactly the fact that the faulty bounds check fails to maintain.
For real libraries, these invariants involve heap-allocated structures, ownership of buffers across calls, and internal state machines, and they must be proved at the level of the compiled binary.
A verifier for verification harnesses therefore needs the agent to discover such invariants and API-level lemmas, to prove them once, and to compose them across calls, rather than to re-establish safety for every call sequence. 
For this reason, it is our position that such realistic agentic verification of a library using verification harnesses will need significant {\em specification inference} (see a recent work from our group \cite{ase26}).
In particular, determining the appropriate level of abstraction at which to formulate these invariants and API-level lemmas requires careful design.
They should be concrete enough to be provable at the machine code level, yet general enough to be composed and reused.
Furthermore, proving these lemmas at the machine code level may be complicated by low-level implementation details.
Therefore, recovering and incorporating the abstractions present in the higher-level source program may be necessary to guide the verification process and make machine code verification more scalable. 
This was most likely not needed for pushing through the proofs that Anthropic and Theorem presented using fuzzing harnesses. 

\smallskip

\textbf{Proving Under LTL Constraints.}
Our verification harness admits only legal clients because its control flow implements the LTL properties of the API, so the proof must also reason about the LTL properties.
All the ordering rules we extract, such as $\varphi_{\mathrm{O1}}$ and $\varphi_{\mathrm{O2}}$, are safety properties, and the harness realizes them through an explicit automaton state such as \texttt{st}.
The loop invariant must therefore relate the automaton state to the library state, for instance, that \texttt{st == OPEN} implies that \texttt{h} points to a live, well-formed handle, and that \texttt{st == NONE} implies that no handle is used.
Beyond crash-freedom, the verifier should also establish that every trace of the harness satisfies the extracted LTL properties, which turns the agreement between the harness and its properties from a checked artifact into a proven fact and removes it from the trusted computing base.
Supporting richer protocols, such as properties that require a resource to be eventually released, would further require reasoning about liveness, for example by reducing it to safety or termination~\cite{reducing-popl17,proving-popl07}.
These capabilities, namely invariant discovery over persistent state, compositional API-level lemmas, and protocol-aware reasoning, are what the released verification harnesses are designed to challenge.

%% file: sections/related-work.tex
\section{Related Work}
\label{sec:related}

Formal verification has long delivered strong guarantees for critical systems, but at the price of substantial expert effort. The seL4 microkernel~\cite{sel4} is the canonical example.
Recent work reduces this effort with LLM agents that construct machine-checked proofs.
\autorocq~\cite{autorocq} applies an agentic approach to generate Rocq proofs for software systems, and OSS-Verify~\cite{anthropic2026formalverification,theorem2026ossverifyproofs} scales the idea to open-source libraries by deriving specifications mechanically from existing fuzz harnesses.
These efforts focus on \emph{how} to prove a given specification. In OSS-Verify, the specification inherits the single call sequence and calling context of the fuzz harness (\autoref{sec:insufficient}).
This direction fits the broader vision of agentic software engineering, in which the focus of programming shifts from scale to trust in what agents produce~\cite{agentic-ai-software-engineers}. A proof is only as trustworthy as the specification it establishes.
Our work addresses the complementary problem of \emph{what} to prove, namely a verification harness that covers all legal usages of an API.
Our use of temporal logic for this purpose relates to verifying temporal properties of programs~\cite{pnueli1977ltl}.
Prior work on liveness and termination~\cite{reducing-popl17,proving-popl07} studies how to \emph{establish} such properties for a given program, e.g., by reducing liveness to safety or to termination.
In contrast, we use LTL properties to \emph{specify} the legal API-usage protocol that constrains the harness, and we leave the proof of the resulting theorem to the entry-point proof pipeline.

In summary, existing verification efforts either demand expert-written specifications or reuse the fuzz harnesses that projects already maintain for continuous fuzzing~\cite{ossfuzz}.
Our vision is to provide verification harnesses that model all legal usages of an API and are constructed automatically from the code and documentation.
With such harnesses, verifying an open-source library becomes as easy as launching a fuzzing campaign, while the guarantee extends from one scenario to every well-behaved client.

%% file: sections/perspective.tex
\section{Perspectives}
\label{sec:perspective}

Formal verification at scale using AI is a topic of high interest. As AI-based formal verification takes centre-stage - the focus shifts to two key aspects.
\begin{itemize}[leftmargin=*]
\item how to verify real software systems using AI agents instead of protocols.

\item more importantly, how to find specifications to capture the operating environments of software being checked, and the properties to check. 
\end{itemize}
In this work, we have taken forward our recent past experience in agentic formal verification of software \cite{autorocq}
to focus on the environment specifications of a software library. We posit that fuzzing harnesses should not be used as environment specifications for formal verification.

We present verification harnesses, which shift a proof about a single usage scenario into a guarantee for every legal client, and an agentic workflow that constructs them automatically.
The harnesses we construct form a set of ready-to-use, library-level verification targets, which we release as open challenges so that future work can focus on \emph{how} to prove them, as well as on richer properties than crash-freedom, proofs that compose across library dependencies, and support for callbacks, global state, and concurrency. 

We also seek to make another observation related to fuzz testing while presenting our findings. Even though the verification harnesses we presented are meant for formal verification - they can be used equally well, for {\em more extensive fuzzing of a software library as well}.  Thus, the fuzzing community could also benefit from our efforts. Indeed in our experiments, we observed new bugs surface when we subjected the open source libraries to testing using the verification harnesses. 

We envision a future in which verifying an open-source library is as routine as fuzzing it.
We recall how  OSS-Fuzz made continuous fuzzing standard practice by asking projects only for a harness. Similarly, moving forward, a software project would contribute a verification harness, agents would prove it, the agent-constructed proof would be validated by a theorem prover. Furthermore, the proof would be re-checked and repaired as the library evolves.
In this setting, the verification harness and its temporal properties become the central, trust-critical artifact, maintained and reviewed alongside the code as the library's API contract.
We invite the verification and AI communities to build on our harnesses and advance agentic verification from proving based on fuzz harnesses, to proving {\em actual} software libraries.

%% file: sections/data-ava.tex
\section*{Data Availability}



We release the verification harnesses as open-source challenges for the verification community.
More information can be found at: 

\noindent
\begin{center}
\textcolor{blue}{\url{https://pre-qed.github.io/}}
\end{center}